\documentclass{article}
\usepackage{spconf,amsmath,graphicx}
\usepackage{booktabs}
\usepackage{enumitem}
\setlist{nosep, leftmargin=14pt}

\usepackage{mwe} 
\usepackage{url}
\usepackage[dvipsnames]{xcolor} 

\title{Toward Contrast-Free Breast MRI: Improving Cancer Detection With Virtual Contrast Enhanced Imaging}
\name{
\begin{tabular}{c}
Hongxi Chen$^{1,2}$ \qquad
Sunwoo Kwak$^{1,2}$ \qquad
Sanyukta Adap$^{1}$ \\
Katerina Dodelzon$^{3}$ \qquad
Paula Feldman$^{2,3}$ \qquad
Mert R. Sabuncu$^{1,2,3}$
\end{tabular}
}

\address{$^{1}$ Cornell University, Ithaca, NY, USA \\
         $^{2}$ Cornell Tech, New York, NY, USA \\
         $^{3}$ Weill Cornell Medicine, New York, NY, USA}
\begin{document}

\maketitle

\begin{abstract}
Although contrast-enhanced (CE) breast MRI is an important tool for cancer detection, intravenous gadolinium administration introduces additional considerations for patients and clinical workflows. AI based virtual contrast enhanced (VCE) MRI could help identify patients for whom acquired CE imaging may be avoidable without compromising diagnostic performance. We developed a framework that combines non contrast (NC) and VCE MRI predictions for cancer detection. On the held-out test set, cancer detection using NC MRI alone achieved an AUROC of 0.865. Combining the predictions from NC and VCE MRI improved AUROC to 0.901. Building on this complementary predictive value, we evaluated patient-level routing to progressively replace acquired CE examinations with the NC plus VCE pathway, prioritizing patients with the most confident NC predictions. This simulation suggests that diagnostic performance can be largely preserved while limiting acquired CE imaging to approximately 40\% of patients. Together, these findings demonstrate the complementary value of VCE MRI and suggest its potential to support selective contrast use in breast cancer screening.
\end{abstract}

\begingroup
\renewcommand{\thefootnote}{}
\footnotetext{%
  \raggedright
  Code available at: \url{github.com/XiGabe/VirtualContrast-BreastMRI}\par
}
\endgroup

\section{Introduction}

Contrast enhanced (CE) magnetic resonance imaging (MRI) plays an important role in breast cancer imaging by providing information about tumor vascularity and lesion characteristics that may not be apparent on non contrast (NC) sequences or other imaging modalities. This information supports cancer detection, lesion characterization, and surgical planning, where accurate delineation of tumor extent is crucial \cite{wekking2023breast}. However, CE MRI requires intravenous administration of gadolinium-based contrast agents, which increases examination time and cost, and may be unsuitable for some patients. Identifying patients who can be adequately evaluated without contrast, while reserving CE imaging for those who need it, could reduce unnecessary contrast administration while preserving diagnostic performance.

Recent advances in artificial intelligence have enabled the synthesis of CE breast MRI from NC acquisitions \cite{ibarra2025comparing, lang2025temporal, liebert2025impact, muller2023using}. However, existing methods have primarily focused on image reconstruction, evaluating synthetic images using image-level similarity metrics rather than clinical utility. Although these studies demonstrate the feasibility of synthesizing contrast enhancement, image similarity does not establish whether clinically relevant information for cancer detection is preserved. Indeed, lesion level evaluations have reported imperfect preservation of enhancement, including poor overlap between synthesized enhancement and corresponding lesions \cite{joshi2026dense}. Moreover, methods developed primarily in malignant cases may not generalize to routine clinical populations, where many examinations contain no suspicious lesions or enhancement. 

Meanwhile, most deep learning approaches to breast cancer detection and lesion characterization rely on acquired CE MRI \cite{DuaYao_ClinRAGGRAPH_MICCAI2026, HamBen_Feature_MICCAI2026, joshi2024leveraging}. Although recent work has explored AI based triage of breast MRI protocols to identify examinations unlikely to contain suspicious findings \cite{nguyen2026adapted}, it remains unclear whether virtual contrast enhanced (VCE) MRI provides diagnostic information beyond the original NC images and whether this information can support patient-level decisions about the need for acquired CE imaging.

In this work, we investigate whether virtual contrast enhanced MRI can extend the diagnostic utility of non contrast breast MRI and enable selective use of acquired contrast enhanced imaging. We hypothesize that VCE MRI captures enhancement related information complementary to NC MRI, improving cancer detection when both are considered jointly. Beyond evaluating image synthesis through its downstream diagnostic value, we assess whether this information can support a contrast sparing clinical pathway. We first compare cancer detection performance using NC, VCE, and acquired CE MRI, evaluating both standalone predictions and the incremental value of combining them. We then simulate selective contrast administration by routing patients to either a contrast free pathway or a standard pathway using acquired CE MRI. We compare random routing with confidence-based routing, which prioritizes patients with more confident NC-MRI predictions for the contrast-free pathway. This framework allows us to quantify the trade-off between reducing acquired contrast use and preserving cancer detection performance, directly testing the potential of virtual contrast enhancement to inform patient level decisions about contrast administration.

\section{Methods}
\subsection{Cohort and Preprocessing}

We used paired fat-suppressed bilateral NC and CE breast MRI from DUKE~\cite{saha2018duke}, MAMAMIA~\cite{garrucho2025mamamia}, and ODELIA~\cite{muller2025odelia}. The first acquired CE phase was used for each examination. Duplicate DUKE records overlapping with MAMAMIA were removed. Each bilateral NC volume was separated into left and right breast volumes using BreastDivider~\cite{rokuss2025breastdivider}. The same crop coordinates were applied to the paired CE volumes to maintain spatial correspondence. Each breast was assigned a binary label (malignant or no finding) based on the source-dataset annotations. Breasts with benign lesions were excluded due to their scarcity. Among the three datasets, ODELIA was the only one that included full examinations without lesions.

Paired NC and CE volumes were normalized using the same
patient-specific intensity parameters derived from the NC volume and resampled to a voxel spacing of $0.8 \times 0.8 \times 2$ mm. Volumes exceeding the in-plane canvas were downscaled
while preserving their aspect ratio, and zero-padding was
applied to obtain volumes of size $S \times 280 \times 336$,
where $S$ denotes the variable number of slices.
The final cohort comprised 2,736 breasts from 1,410 patients.
Patients were split at the patient level into training (968 patients,
1,888 breasts), validation (197 patients, 375 breasts), and test
(245 patients, 473 breasts) sets. The test set contained 192
malignant and 281 no-finding breasts.

\subsection{Virtual Contrast Enhanced MRI Generation}

We trained a 3D U-Net~\cite{cicek2016unet} to generate VCE MRI from the corresponding NC acquisition. Rather than directly predicting the post-contrast image, the generator was trained to predict the contrast enhancement. Given an NC image $x$ and its paired acquired CE image $y$, the enhancement is defined as $d=y-x$. The generator predicts this residual as $\hat{d}=G(x)$, and the VCE image is reconstructed as
\begin{equation}
\hat{y}=x+\hat{d}.
\end{equation}

The generator was optimized using a composite objective designed to promote accurate image reconstruction while emphasizing contrast-enhancing regions and penalizing both missed and excessive enhancement:
\begin{equation}
\mathcal{L}_G =
\mathcal{L}_{\mathrm{rec}}
+\alpha\mathcal{L}_{\mathrm{enh}}
+\beta\mathcal{L}_{\mathrm{miss}}
+\gamma\mathcal{L}_{\mathrm{excess}}.
\label{eq:generator}
\end{equation}
The reconstruction loss $\mathcal{L}_{\mathrm{rec}}$ combines an L1 loss within the breast region with a lower-weighted penalty outside it. The enhancement-weighted loss \(\mathcal{L}_{\mathrm{enh}}\) assigns greater weight to reconstruction errors in regions with larger positive intensity differences ($d=y-x$), while $\mathcal{L}_{\mathrm{miss}}$ penalizes underestimated enhancement and $\mathcal{L}_{\mathrm{excess}}$ penalizes overestimated enhancement.

\subsection{Cancer Detection}

We trained three independent breast cancer detection classifiers based on the Medical Slice Transformer (MST)~\cite{muller2025mst}, each trained on a different MRI representation: non contrast (NC), acquired contrast enhanced (CE), or virtual contrast enhanced (VCE) MRI. We refer to these as NC Classifier, CE Classifier, and VCE Classifier.

Given a breast MRI volume of size $S\times280\times336$, each slice is processed independently using a DINOv2 ViT-S/14 encoder~\cite{oquab2024dinov2}, with the input replicated across three channels. The encoder produces a 384-dimensional representation for each slice. The resulting sequence of slice representations is aggregated using a single-layer transformer, followed by a linear classification head to produce a breast-level cancer prediction. The encoder and aggregation transformer were fine-tuned end-to-end.

We evaluated the classifiers in the following settings: (1) NC MRI using the NC Classifier; (2) VCE MRI using the CE Classifier; (3) VCE MRI using the VCE Classifier; (4) acquired CE MRI using the CE Classifier; (5–6) fusion of NC predictions with VCE predictions from the CE and VCE Classifiers, respectively; and (7) fusion of NC and acquired CE predictions as a reference. For settings involving two classifiers, we averaged their logits and applied a sigmoid function to obtain the final breast-level cancer probability.

\subsection{Implementation Details}
The generator consisted of a five-stage 3D U-Net with instance normalization. Training was performed for 10 epochs using AdamW with cosine learning-rate decay from \(10^{-4}\) to \(10^{-5}\). The generator objective combined reconstruction, enhancement-weighted, missed-enhancement, and excess-enhancement losses, weighted by \(\alpha=2\), \(\beta=1\), and \(\gamma=0.03\), respectively. We used \(64\times256\times256\) 3D patches, with half of the training patches sampled from regions exhibiting stronger contrast enhancement. At inference, VCE images were generated using 50\%-overlapping sliding windows and reconstructed to the full breast volume.
All three classifiers were trained independently using binary cross-entropy loss and dataset-by-label balanced sampling. Optimization was performed following Medical Slice Transformer's hyperparameters. Training was performed for up to 60 epochs, with the checkpoint achieving the highest validation AUROC selected for evaluation.

\begin{figure}[t]
    \centering
    \includegraphics[width=\linewidth]
    {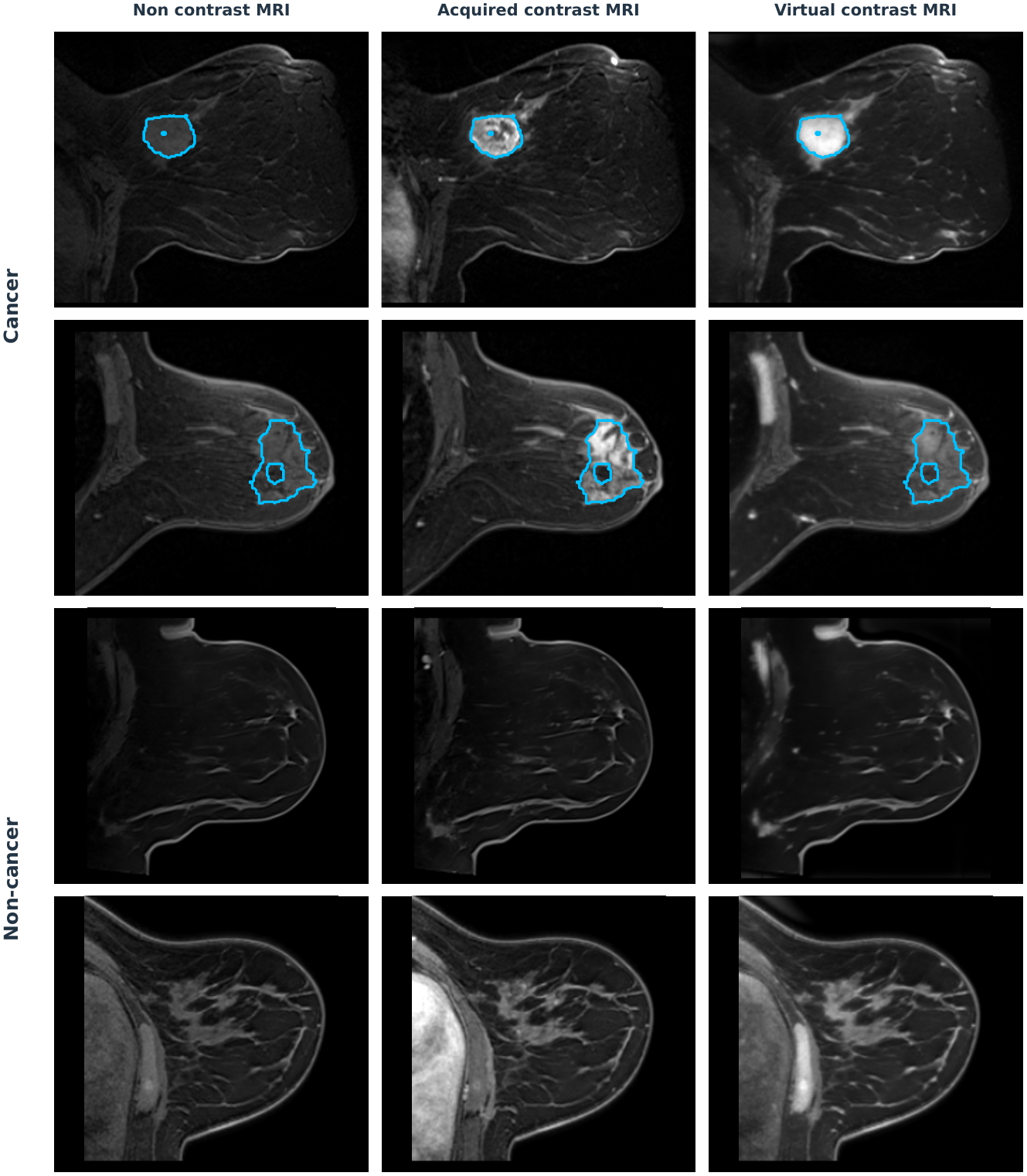}
    \caption{Qualitative examples of virtual contrast enhanced MRI. Blue outlines indicate annotated lesions.}
    \label{fig:qualitative_synthesis}
\end{figure}

\section{experiments and results}

\subsection{Cancer Detection}

We evaluated whether VCE MRI provides complementary information for breast cancer detection beyond NC MRI. On the held-out test set, we compared the following settings: (1) NC MRI using the NC Classifier; (2–3) VCE MRI using the CE and VCE Classifiers, respectively; (4) acquired CE MRI using the CE Classifier; (5–6) fusion of NC predictions with VCE predictions from the CE and VCE Classifiers, respectively; and (7) fusion of NC and acquired CE predictions as a reference. Fusion predictions were obtained by averaging classifier logits. Performance was assessed using AUROC and AUPRC across three classifier repeats. Qualitative examples are shown in Fig.~\ref{fig:qualitative_synthesis}, and quantitative results in Table~\ref{tab:cancer_detection}.

VCE MRI achieved standalone discrimination comparable to NC MRI (AUROC $0.866$ vs. $0.865$), with higher performance when evaluated using a classifier trained on virtual images (3). Combining predictions from NC MRI and VCE MRI further improved discrimination, reaching AUROCs of $0.901$ and $0.896$ with the acquired contrast trained (5) and virtual contrast trained (6) classifiers, respectively, compared with $0.865$ for NC MRI alone (1). These gains support the complementary predictive value of VCE MRI. Notably, the improvement in cancer detection when combining NC and VCE MRI suggests that virtual contrast enhancement provides diagnostically useful information that is not fully captured by the original NC images.

Fusion gains over NC alone were consistent across DUKE, MAMAMIA, and ODELIA. AUROC improved by $0.027$, $0.042$, and $0.046$, respectively, using the CE Classifier on VCE MRI, with positive gains also observed using the VCE Classifier. These consistent improvements suggest that VCE MRI provides complementary predictive information across datasets. However, the lower overall performance on ODELIA highlights the challenges of translating these gains to more representative clinical populations, where examinations may show no suspicious lesions or enhancement. Evaluating models in such populations is essential to establish their generalizability and their potential to guide selective contrast administration in clinical practice.
\begin{table*}[t]
\centering
\caption{Cancer detection performance on the held-out test set, overall and by source dataset. Overall AUROC and AUPRC are reported as mean $\pm$ SD across three classifier repeats. }
\label{tab:cancer_detection}

\small
\setlength{\tabcolsep}{8pt}
\renewcommand{\arraystretch}{1.12}

\begin{tabular}{@{}clccccc@{}}
\toprule
\textbf{Exp.} &
\textbf{Input / strategy} &
\multicolumn{2}{c}{\textbf{Held-out test set}} &
\multicolumn{3}{c}{\textbf{Dataset-specific AUROC}} \\
\cmidrule(lr){3-4}
\cmidrule(lr){5-7}
& &
\textbf{AUROC} &
\textbf{AUPRC} &
\textbf{DUKE} &
\textbf{MAMAMIA} &
\textbf{ODELIA} \\
\midrule

1 & NC MRI (NC Classifier)
& $0.865\pm0.007$
& $0.831\pm0.020$
& $0.861\pm0.005$
& $0.897\pm0.017$
& $0.795\pm0.012$ \\

2 & VCE MRI (CE Classifier)
& $0.866\pm0.007$
& $0.841\pm0.005$
& $0.864\pm0.014$
& $0.891\pm0.011$
& $0.799\pm0.016$ \\

3 & VCE MRI (VCE Classifier)
& $0.883\pm0.010$
& $0.854\pm0.022$
& $0.882\pm0.003$
& $0.908\pm0.007$
& $0.833\pm0.029$ \\

4 & CE MRI (CE Classifier)
& $0.945\pm0.003$
& $0.927\pm0.007$
& $0.950\pm0.013$
& $0.966\pm0.001$
& $0.900\pm0.010$ \\

\midrule

5 & NC + VCE (NC + CE Classifiers)
& $0.901\pm0.005$
& $0.874\pm0.005$
& $0.888\pm0.008$
& $0.939\pm0.012$
& $0.841\pm0.003$ \\

6 & NC + VCE (NC + VCE Classifiers)
& $0.896\pm0.002$
& $0.870\pm0.004$
& $0.890\pm0.008$
& $0.927\pm0.007$
& $0.837\pm0.021$ \\

7 & NC + CE (NC + CE Classifiers)
& $0.943\pm0.010$
& $0.924\pm0.012$
& $0.942\pm0.012$
& $0.972\pm0.007$
& $0.888\pm0.021$ \\

\bottomrule
\end{tabular}
\end{table*}

\subsection{Contrast Sparing Simulation}

We simulated selective replacement of acquired CE MRI with a contrast-sparing pathway combining NC and VCE predictions in the held-out test set. As a reference, patients were randomly assigned to the contrast-sparing pathway over 2,000 permutations, progressively increasing the replacement rate. Breast level AUROC and AUPRC were averaged across three classifier repeats using identical patient assignments; 95\% randomization intervals quantified variability across permutations.

For confidence-based routing, patients were ranked by NC prediction confidence, defined as $\max(p,1-p)$ and averaged across the three repeats for each breast. The minimum confidence across a patient's included breasts determined the patient-level ranking, ensuring all breasts from each patient followed the same pathway. Patients were progressively routed to the contrast-sparing pathway in descending confidence order, and performance was compared with the random-replacement intervals at matched replacement rates.

\begin{figure}[t]
    \centering
    \includegraphics[width=\linewidth]
    {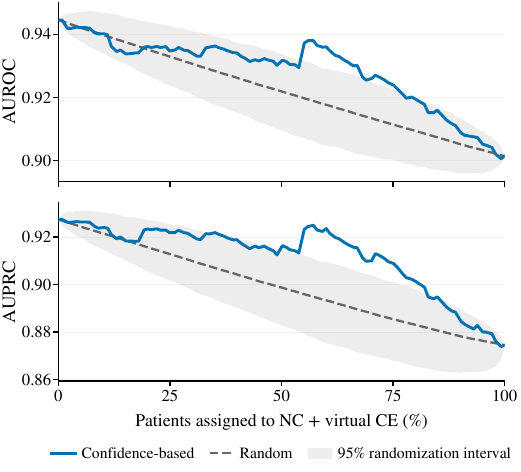}
    \caption{Selective contrast-sparing.
    AUROC and AUPRC versus the proportion of patients assigned
    to the NC plus VCE pathway. Shaded regions indicate
    95\% randomization intervals under random patient assignment.}
    \label{fig:selective_routing}
\end{figure}

As shown in Fig.~\ref{fig:selective_routing}, confidence-based routing maintained higher AUROC and AUPRC than random replacement across a broad range of replacement rates, with slower performance decline as contrast use decreased. These findings suggest that NC prediction confidence may help prioritize patients for contrast-sparing imaging.

\section{Discussion and conclusions}

While VCE imaging is not yet a substitute for acquired contrast, adding virtual contrast to the NC examination meaningfully narrowed the gap. The combined approach improved AUROC, closing nearly half the diagnostic gap between NC and CE imaging. This benefit was consistent across all three independent datasets. Notably, VCE alone did not significantly outperform NC MRI; its value lies in complementing, not replacing, the NC examination.
Traditionally reserved for women at high risk, MRI screening is increasingly extended to a broader population, including women with dense breasts, who are often at lower risk of breast cancer. In these patients, the balance between the benefit of contrast and its burden shifts: most examinations will be negative, yet each requires gadolinium administration, often annually over many years. Because NC sequences are acquired before injection, routing could be applied at the time of the scan: patients with high confidence NC predictions would complete a shorter, contrast free examination. In our simulation, this approach largely preserved diagnostic performance while limiting acquired CE imaging to approximately 40\% of patients. In practice, it could reduce contrast exposure and cost, shorten examinations, and expand MRI capacity as screening indications grow. Acquired CE MRI would remain the standard for lesion characterization, assessment of disease extent, and biopsy guidance.
This study has several limitations. The datasets are drawn predominantly from cancer-enriched and preoperative cohorts rather than screening populations, and overall sample size was limited. Future work will validate this approach in larger and more diverse screening cohorts that include benign findings, evaluate radiologist performance with synthetic images, and assess prospective, real-time routing.

\subsection*{Compliance with Ethical Standards}
This study uses retrospective publicly available datasets. No additional ethical approval was required.

\subsection*{Conflict of Interest}

The authors declare that they have no conflicts of interest related to this work.

\subsection*{Acknowledgments}
This work was supported in part by the National
Institutes of Health (NIH) under grants
R01HL174863 and U54DK144866.

\bibliographystyle{IEEEbib}
\bibliography{refs}
\end{document}